\documentclass[epsfig,epsf]{article}
\usepackage{amsmath,amssymb,graphicx,multirow,subfigure,oldgerm,euscript,mathrsfs}
\begin{document}

\begin{flushright}
\begin{tabular}{l}
  PCCF-RI-0414 \\
  ${\rm ECT}^{*}$-04-10 \\
\end{tabular}  
\end{flushright}

\begin{center}

{{\Large \bf Direct CP Violation in $B$ decays with  
 $\boldsymbol{\rho^{0}-\omega}$ Mixing}} \\

\vspace{4mm}

Ziad J. Ajaltouni \footnote{Collaborating authors~: C.Rimbault, O.Leitner, P.Perret, A.W.Thomas} \\
Laboratoire de Physique Corpusculaire \\
Universit\'e Blaise Pascal CNRS/IN2P3 \\
F63177 AUBIERE CEDEX FRANCE \\
\end{center}

\begin{abstract}
A complete study of the processes $ B \to {\pi}^+ {\pi}^- V \ \ (V = 1^{--})$
is performed both in the framework of the helicity formalism and the effective lagrangian approach.
Emphasis is put on the factorization hypothesis and the importance of the ${{\rho}^{0}}-{\omega}$ 
mixing in enhancing the direct $CP$ violation. New results involving some branching ratios and the ratio
of the Penguin/Tree amplitude are given in details.
\end{abstract}

%

\section{Physical Motivations for the channels $B \to V_1 V_2 $ and their Decay Kinematics}
%

In the framework of the LHCb experiment devoted to the search for $CP$ violation
and rare $B$ decays, special care is given to the $B$ decays into two 
vector mesons, $B \to  V_1 V_2, \ \ V_i  =  \ 1^{--}$. 

(i) $B$ decays being governed by weak interactions, the vector-mesons 
are {\it polarized} and their final states 
have specific angular distributions; which allows one to cross-check 
the Standard Model (SM) predictions and to 
perform tests of models {\it beyond} the SM. \\
(ii) In the special case of two neutral vector mesons with ${\bar V}^0 = V^0 \ $; 
orbital angular momentum 
$ \ell$, total spin $S$ and $CP$ eigenvalues are related by 
the following relations~:  
$ \ell = \ S \ = \ 0,1,2   \  \Longrightarrow  CP \ = \ {(-1)}^{\ell}\ $ , 
 which implies a {\it mixing of different $CP$ eigenstates}, proving a  
$CP$ non-conservation process. 
According to Dunietz {\it et al}~\cite{Dunietzetal}, tests of $CP$ violation 
in a model independent way can 
be performed and severe constraints on models beyond the SM can be set.
Because the $B$ meson has spin $0$, the final two vector 
mesons, $V_1$ and $ V_2$, have the same helicity 
${\lambda}_1 = {\lambda}_2 = -1,0, +1,$ and their 
angular distribution is isotropic in the $B$ rest frame.
Let $H_w$ be the weak Hamiltonian describing the $B$ decays. Any transition
amplitude between the initial and final states will have the following form:
\begin{equation}\label{eq1}
 H_{\lambda} = \langle V_1{(\lambda)} V_2{(\lambda)}|H_w|B \rangle\ 
\end{equation} 
where the common helicity is ${\lambda} = -1,0, +1$.  
Then, each vector meson $V_i$ will decay into two 
pseudo-scalar mesons, $a_i \  \mathrm{and} \ b_i$ ;
$a_i ( b_i)$ can be either a pion or a kaon which angular
distributions  depend on $V_i$ polarization.

The helicity frame of a vector-meson $V_i$ is defined in the $B$ 
rest frame  such that the direction of the Z-axis is given by its momentum 
${\vec{p}_i}$. Schematically, the whole process  gets the form:
\begin{eqnarray*}
 B     \longrightarrow     V_1  +   V_2  \longrightarrow   (a_1 + b_1)  +
 (a_2 + b_2)\ .
\end{eqnarray*}      
The corresponding decay amplitude, $M_{\lambda}\bigl(B \rightarrow \sum_{i=1}^2
(a_i+b_i)\bigr)$, is factorized according to the relation,
\begin{equation}\label{eq2}
 M_{\lambda}\bigl(B \rightarrow \sum_{i=1}^2 (a_i+b_i)\bigr) = 
  H_{\lambda}(B \rightarrow V_1 +V_2) \times \prod_{i=1}^2 A_i(V_i \rightarrow 
a_i + b_i)\ ,    
\end{equation}
where the amplitudes $A_i(V_i \rightarrow a_i + b_i)$ are 
related to the decay of the resonances
$V_i$. For a given value of ${\lambda}$ and a well defined final state,  
amplitudes $A_i(V_i \rightarrow a_i + b_i)$ are given, according to the Wigner-Eckart theorem,
by the following expressions:
%
\begin{equation}\label{eq3}
A_1(V_1 \rightarrow a_1 + b_1)  =    c_1 D^1_{\lambda m_1}(0,\theta_1,0) \ \  \mathrm{and} \ \ 
A_2(V_2 \rightarrow a_2 + b_2)  =   c_2 D^1_{\lambda m_2}(\phi,\theta_2,-\phi)\ .  
\end{equation}
 
 In Eq.~(\ref{eq3}), the $c_1$
and $c_2$ coefficients represent respectively
the {\it dynamical  decay parameters} of the $V_1$
and $V_2$ resonances. 
The term $D^1_{\lambda m_i}(\phi_i,\theta_i,-\phi_i)$ is the 
Wigner rotation matrix element for a spin-1 particle and we
let $\lambda{(a_i)}$ and $\lambda{(b_i)}$ be the respective helicities of the final particles $a_i$ and $b_i$
in the $V_i$ rest frame. $\theta_1$  is the polar angle of
$a_1$ in the $V_1$ helicity frame. 
The decay plane of $V_1$ is identified with  
the (X-Z) plane and consequently the azimuthal angle 
$\phi_1$ is set to $0$. Similarly
$\theta_2$ and $\phi$ are respectively the polar and azimuthal angles of
particle $a_2$ in the $V_2$ helicity frame. Finally, the coefficients $m_i$
are defined as $ \  m_i=\lambda(a_i)-\lambda(b_i)$.
%

\section {Decay Dynamics and Basis for Simulations } 

The most general form of the decay amplitude ${\cal M}\bigl(B 
\rightarrow \sum_{i=1}^2
(a_i + b_i)\bigr)$ is a {\it linear superposition} 
of the previous amplitudes $M_{\lambda}\bigl(B \rightarrow \sum_{i=1}^2 (a_i
+b_i)\bigr)$ denoted by:
\begin{equation}\label{eq4}
 {\cal M}\bigl(B \rightarrow \sum_{i=1}^2 (a_i +b_i)\bigr)= 
  \sum_{\lambda} M_{\lambda}\bigl(B \rightarrow 
\sum_{i=1}^2 (a_i +b_i)\bigr)\ . 
\end{equation} 
 
The decay width $\Gamma{(B \rightarrow V_1 V_2)}$ can be
computed by taking the square of the modulus,  
$|{\cal M}\bigl(B \rightarrow \sum_{i=1}^2 (a_i +b_i)\bigr)|$,
which involves the three kinematic parameters,  
$\theta_1, \theta_2$ and  $\phi$. This leads to the following
general expression: 
\begin{equation}\label{eq5}
d^3{\Gamma}(B \rightarrow V_1 V_2)  \propto  \Bigl|\sum_{\lambda}
M_{\lambda} \bigl( B \rightarrow
\sum_{i=1}^2 (a_i +b_i)\bigr)\Bigr|^2  =  \sum_{\lambda,\lambda'}
 h_{\lambda, \lambda'} F_{\lambda,\lambda'}(\theta_1) G_{\lambda,\lambda'}
 (\theta_2, \phi)\ ,
\end{equation} 
which gives rise to  {\bf three density-matrices}
$ \ h_{\lambda  \lambda'}, F_{\lambda  \lambda'}(\theta_1)$ and 
$G_{\lambda  \lambda'}(\theta_2, \phi)$ ~:
(i) The factor $h_{\lambda  \lambda'}= H_{\lambda} H^{*}_{\lambda'}$ is an 
element of the density-matrix related to the $B$ decay;
 (ii)  $F_{\lambda  \lambda'}(\theta_1)$ represents the density-matrix of 
the decay $V_1 \rightarrow a_1 + b_1$ and (iii)
$G_{\lambda  \lambda'}(\theta_2, \phi)$ represents the
density-matrix of the decay $V_2 \rightarrow a_2+b_2$.
\par
 
\noindent The analytic expression in Eq.~(\ref{eq5}) exhibits 
a {\bf very general form}.  It depends on neither 
the specific nature of the intermediate resonances nor their 
decay modes (except for the spin of the final particles). 

The previous calculations are illustrated by the reaction $B^0 \rightarrow
K^{* 0} {\rho}^0$ where $K^{* 0} \rightarrow K^+ {\pi}^-$ and 
${\rho}^0 \rightarrow {\pi}^+ {\pi}^{-}$. In this channel, since all the final
particles have spin zero,  the coefficients  $m_1$ and $m_2$, defined
previously, are equal to zero. The three-fold differential width has the following form~:
\begin{multline}\label{eq6}
\frac{d^3\Gamma(B \rightarrow V_1 V_2)}{d(\cos\theta_1) 
d(\cos\theta_2) d\phi}  \propto 
\bigl(h_{++} + h_{--}\bigr){{\sin}^2{\theta_1}{\sin}^2{\theta_2}}/4 +
{h_{00}{\cos}^2{\theta_1}{\cos}^2{\theta_2}} \\
+ \Bigl\{\mathscr{R}\!e{(h_{+0})}{\cos{\phi}} - \mathscr{I}\!m{(h_{+0})}{\sin{\phi}} + \mathscr{R}\!e{(h_{0-})}{\cos{\phi}} -
\mathscr{I}\!m{(h_{0-})}{\sin{\phi}}\Bigr\}{{\sin{2\theta_1}}{\sin{2\theta_2}}}/4  \\
+ \Bigl\{\mathscr{R}\!e{(h_{+-})}{\cos{2\phi}} - \mathscr{I}\!m{(h_{+-})}{\sin{2\phi}}\Bigr\}
{{\sin}^2{\theta_1}{\sin}^2{\theta_2}}/2\ .
\end{multline}
It is worth noticing that the  expression in Eq.~(\ref{eq6}) is {\it
completely symmetric} in ${\theta}_1$ and ${\theta}_2$ and consequently the
angular distribution of $a_1$ in the $V_1$ frame is {\it identical} to that of
$a_2$ in the $V_2$ frame. From Eq.~(\ref{eq6}) 
the normalized probability distribution functions (pdf) of ${\theta}_1$, 
${\theta}_2$ and $\phi$ can be derived and one finds~:
%
\begin{equation}\label{eq7}  
 f{({\cos}{\theta}_{1,2})} \  =  {(3h_{00}-1)}{{\cos}^2{\theta}_{1,2}}  +  {(1-
 h_{00})}\ ,   
 \ \ g{(\phi)} \  =  1 + 2 \; \mathscr{R}\!e{(h_{+-})}{\cos{2\phi}}  -  2 \; 
\mathscr{I}\!m{(h_{+-})}{\sin{2\phi}}\ . 
\end{equation}
%
A practical way to compute the matrix elements is to use the Effective Hamiltonian approach
based on the general hamiltonian~:

\begin{equation}\label{eq8}
{\cal H}_{eff}=\frac {G_{F}}{\sqrt 2} \sum_{i} V_{CKM} C_{i}(\mu)O_i(\mu)\ ,
\end{equation}

\noindent where $G_{F}$ is the Fermi constant, $V_{CKM}$ is the 
CKM matrix element,
$C_{i}(\mu)$ are
the Wilson Coefficients (W.C.),
$O_i(\mu)$ are the operators associated to the {\bf tree, QCD-penguin and 
EW-penguin} diagrams and, finally $\mu$ is the renormalization energy scale taken equal to $m_B$.
 
 \par
 Then, applying the Operator Product Expansion (OPE) method pioneered by Wilson, the 
W.C. $C_i$ are calculated perturbatively at the Next to Leading Order (NLO) for an energy scale 
$ \ge m_B$ \cite{Burasetal}. The non-perturbative effects which are   
related to the operators $O_i$ and representing physical processes at an energy $\le m_B$ 
are introduced through a set of form factors. The latters  are explicitly
computed in the framework of the pioneering BSW models \cite{BSW}.
However some free parameters remain like~: 
(i) the ratio ${q^2}/{{m_b}^2}$ where ${q^2}$ is the squared invariant mass 
of the gluon appearing in the penguin diagrams and 
(ii) the effective number of colors ${N_c}^{eff}$. 

\subsubsection*{Final State Interactions and  ${\rho}^0 - {\omega}$  Mixing}

Hadrons produced from $B$ decays are scattered again by their mutual 
strong interactions, which could modify completely their final wave-function. 
Computations of the branching ratios $B \to {\rm {hadrons}}$ must take 
account of the
final state interactions (FSI) \cite{Quinn} which are generally divided 
into  two regimes~:
 {\it perturbative} and {\it  non-perturbative}. These two aspects have been already mentioned
above in the framework of the OPE method. However an
important question arises: how to deal with the FSI in a simple and 
practical way in order to perform realistic and rigorous simulations? \\
\noindent The method which has been followed for the computations is largely 
developed in \cite{Ziadetal} and \cite{Rimbault}  and it is based on the hypothesis 
of {\bf Naive Factorization}, which can be summarized as
follows~: \\
 $\star$ In the Feynman diagrams describing the $B$ decays into hadrons like 
{\it tree or penguin} diagrams, the soft gluons exchanged among the quark lines 
are {\it neglected}. \\
 $\star$ The color number $N_c$ is no longer fixed and equal to 3. It is modified according 
to the relation~: \\
$\frac{1}{(N_{c}^{eff})} = \frac{1}{3} + \xi\ \ $ where $\xi$ is an operator representing the
 non-perturbative effects. \\
\noindent $\star$ The QCD-penguin diagram introduces an intrinsic phase-shift, ${\delta}_{P/T} \ $,
by comparison with the tree one (BSS mechanism \cite{BSS}).
  Thus, the total amplitude gets an {\it absorptive part}, which is an illustration of the
 FSI in the perturbative regime. \\
$\star$  Another important effect which appears in the channels 
$ B \to {\pi}^+ {\pi}^- V$ is the {\bf ${{\rho}^0}- {\omega}$ mixing}, 
which is an unavoidable quantum process. Indeed, the tree amplitude $A^T$ and the penguin one,
$A^P$,  are modified according to the following relations~:    
\begin{equation}\label{eq10}
\langle K^{*} \pi^{-} \pi^{+}|H^{T}|B  \rangle 
=
\frac{g_{\rho}}{s_{\rho}s_{\omega}}
 \tilde{\Pi}_{\rho \omega}t_{\omega} +
\frac{g_{\rho}}{s_{\rho}}t_{\rho}\  \ \ ,  \ \ 
\langle  K^{*} \pi^{-} \pi^{+}|H^{P}|B  \rangle 
=
\frac{g_{\rho}}{s_{\rho}s_{\omega}} 
\tilde{\Pi}_{\rho \omega}p_{\omega} +\frac{g_{\rho}}{s_{\rho}}p_{\rho}\ 
\end{equation}
Here $t_{V} \; (V=\rho \;{\rm  or} \; \omega)  \  \mathrm{and}  \ p_{V}$ are respectively 
the tree  and penguin amplitudes for 
producing a vector meson $V \ $, $g_{\rho}$ is the 
coupling for $\rho^{0} \rightarrow \pi^{+}\pi^{-} \ $,
$\tilde{\Pi}_{\rho \omega}$ is the effective $\rho^{0}-\omega$ mixing
amplitude and $s_{V}$ is the inverse 
propagator of the vector meson $V \ ,  
\  \ s_{V}=s-m_{V}^{2}+im_{V}\Gamma_{V} \  \mathrm{where} \  
\sqrt s$ is the invariant mass of the $\pi^{+}\pi^{-}$ pair. \\  
\noindent The ratio ${A^P}/{A^T}$, which is a complex number, gets the final 
expression~: 
\begin{equation}\label{eq11}
 re^{i \delta} e^{i \phi}= \frac{ \tilde {\Pi}_{\rho \omega}p_{\omega}+
s_{\omega}p_{\rho}}{\tilde 
{\Pi}_{\rho \omega} t_{\omega} + s_{\omega}t_{\rho}}\ , 
\end{equation}
\noindent where $\delta$ is the {\it total strong phase} arising both from the 
${{\rho}^0}- {\omega}$ resonance mixing and
the penguin diagram quark loop, and $\phi$ is the weak angle 
resulting from the CKM matrix elements.  
\section {Main Results and Comparison with Recent Experimental Data}

Owing to the presence of resonances with large widthes, the mass of each resonance  
is generated according to a {\it relativistic Breit-Wigner} distribution~:
 \begin{eqnarray*}
\frac{d\sigma}{dM^2}  \propto   \frac{\Gamma_R M_R}{{(M^2-{M^2_R})}^2 +
  {(\Gamma_R M_R)}^2}\ ,
\end{eqnarray*}
%
$M_R$ and $\Gamma_R$ being respectively the mass and the width of the vector meson. \\
 Then, combining both the Wilson Coefficients and the BSW formalism and including the 
${{\rho}^0}-{\omega}$ mixing, 
the helicity amplitude (computed in the $B$ meson rest-frame) is given by the following expression~:
\begin{multline}\label{eq12} 
H_{\lambda}\bigl(B \rightarrow \rho^{0}(\omega) V_2 \bigr) = 
iB^{\rho}_\lambda(V_{ub}V_{us}^{*}c_{t_1}^{\rho}-
V_{tb}V_{ts}^{*}c_{p_1}^{\rho})+
iC^{\rho}_\lambda(V_{ub}V_{us}^{*}c_{t_2}^{\rho}-
V_{tb}V_{ts}^{*}c_{p_2}^{\rho}) + \\ 
\frac{\tilde{\Pi}_{\rho
  \omega}}{(s_{\rho}-m_{\omega}^{2})+im_{\omega}\Gamma_{\omega}}
\Bigl[ iB^{\omega}_\lambda(V_{ub}V_{us}^{*}c_{t_1}^{\omega}-
V_{tb}V_{ts}^{*}c_{p_1}^{\omega})+
iC^{\omega}_\lambda(V_{ub}V_{us}^{*}c_{t_2}^{\omega}-
V_{tb}V_{ts}^{*}c_{p_2}^{\omega})\Bigr]
\ ,
\end{multline} \\
where the terms $B_\lambda^{V_i}$ and $C_\lambda^{V_i}$ are combinations of 
different form factors. Their
explicit expressions, corresponding to the helicity values 
(${\lambda} = \  -1,0,+1$), are given in  Ref.~\cite{Ziadetal}. 
This expression allows to deduce  the dynamical 
density-matrix elements $ \ h_{{\lambda}  {\lambda}'}\ $ given by~:
\begin{eqnarray*}
h_{{\lambda}  {\lambda}'} = H_{\lambda}\bigl(B \rightarrow \rho^{0}
(\omega) V_2 \bigr) H^{*}_{\lambda'}\bigl(B \rightarrow \rho^{0}
(\omega) V_2 \bigr)\ . 
\end{eqnarray*}
Because of the {\it hermiticity} of the DM, only six elements need to 
be calculated. The main results are~: \\
 1) The matrix elements $ \ h_{i j}$ depend essentially on the {\bf masses} of the resonances.
 Their spectrum of is too wide because of the resonance widths, especially the ${\rho}^0$
 width  ${\Gamma}_{\rho} = \ 150  \ {\rm MeV/c^2}. $ \\ 
 2) The  longitudinal polarization, $h_{00} = \  {|H_0|}^2 \ ,$  is 
{\bf largely dominant}. 
  In the case of $B^0 \to \rho^0 (\omega) K^{* 0}$,   the mean value of 
$h_{00}$ is $\approx 87\%$     
while for $B^+ \to \rho^0 (\omega) \rho^{+}$, its  mean value is 
$\approx 90\% \ $ (Fig.1). 
These results have been confirmed recently by both BaBar~\cite{Babar} and 
Belle collaborations~\cite{Belle}. \\
3) The matrix element $h_{--} = {|H_{-1}|}^2$  is very tiny,  $ \ \leq \  0.5\%$. \\
4) The  non-diagonal matrix elements $h_{i j}$ are mainly characterized by~: 
 (i) The {\bf smallness} of both their real and imaginary parts.  
 (ii) The ratio $ \ {\mathscr{I}\!m}/{\mathscr{R}\!e}   \ \approx  0.001 \to  0.1$. 
 (iii) In the special case of $B^+ \to \rho^0 (\omega) \rho^{+}, \  
{\rm \mathscr{I}\!m {(h_{i j})}} \  \approx 0.0 \ $. 
\noindent Our conclusion is that there is a kind of  
{\it universal behavior} of the density-matrix elements,
 whatever the decay 
$B \to {\pi}^+ {\pi}^- V$ is 
($V = \  K^{*0}, K^{*{\pm}} \ , \ {\rho}^{\pm}$). \\
$\bullet$ Consequences on the Angular Distributions~: \\
In the helicity frame of each vector-meson $V_i \ $, the angular distributions 
given above (see Eq.~(7)) become simplified~: 
because of the small value of $\langle h_{+-} \rangle$,
the azimuthal angle distribution $g(\phi)$  is rather {\bf flat}.
In the expression of $f{({\cos}{\theta}_{1,2})} \ $, the longitudinal part $h_{00}$ being largely 
dominant, the polar angle distribution  is $ \approx  {{\cos}^2}{\theta}$.  
\subsubsection*{Branching Ratios and Asymmetries}

$\star$ The energy $E_i$ and  the momentum $p_i$ of each vector meson vary 
significantly according to the generated event. So, the width of each channel 
is computed by Monte-Carlo methods from the fundamental relation~:
\begin{equation}\label{eq13}
 d{\Gamma}(B \rightarrow V_1 V_2) \ = \  {\frac{1}{8{\pi}^2 M}} \ 
{|{\cal M}(B \rightarrow V_1 V_2)|}^2 \ {\frac{d^3{\vec p_1}}{2E_1}} \ 
{\frac{d^3{\vec p_2}}{2E_2}}
 \ {\delta}^4{(P- p_1 - p_2)}\ 
\end{equation}
from which the specific branching ratios are deduced. \\
\noindent $\star$ For a fixed value of ${q^2}/{{m_b}^2}$, the BRs depend  
strongly on the Form Factor models. 
They could vary  up to a {\bf factor 2}.

\noindent $\star$ The relative difference between two conjugate branching ratios, 
 $ Br{(B \rightarrow f)}$ and  
$ Br{({\bar B} \rightarrow {\bar f})}$, 
is almost  {\it independent} of the form-factor models. The {\it global asymmetry} defined by~: 

$$ {\cal A}_{CP} = \frac{Br{(B \rightarrow f)}- Br{(\bar B \rightarrow \bar f)}}{Br{(B \rightarrow f)} +
Br{(\bar B \rightarrow \bar f)}} $$ \\
is usually $ \ \le  2\% $ \\
$\star$ However, an interesting effect is found in the variation of the 
{\bf differential asymmetry} with respect to 
the $\pi \pi$ invariant mass. This parameter defined as~:
\begin{eqnarray*}          
 a_{CP}(m) =   \frac{{\Gamma}_m{(B \rightarrow f)} - {\bar \Gamma}_m{(\bar B
     \rightarrow {\bar f})}}{{\Gamma}_m{(B \rightarrow f)} +
 {\bar \Gamma}_m{(\bar B \rightarrow {\bar f})}} 
\end{eqnarray*}    
 is {\bf amplified} in the in the vicinity of the $\omega$ resonance mass
($\pm 20 \ {\rm MeV/}c^2$ around $ M_{\omega} = \ 782 {\rm MeV/}c^2$). $a_{CP}(m)$ is $\approx  15\%$
in the case of $ B^0 \to  K^{*0} {\rho}^0 {(\omega)} \ $
and equal to $24\%$ in the channel $B^{\pm} \to  {\rho}^{\pm} {\rho}^0 {(\omega)}$.
\par
This kind of asymmetry is almost independent of the form factor models. It is worth noticing that this novel
effect has been predicted analytically in the channel $B \to V P  \to {\pi}^+ {\pi}^- {\pi}$
by Leitner et al \cite{Leitner} and its only explanation is the {\it mixing} process of the two vector-mesons
${\rho}^0 \ \mathrm{and} \  {\omega}$. 
 
 \subsubsection*{Ratio Penguin/Tree} 

The ratio Penguin/Tree is given by the following relation derived from equation (11)~:
  
$$ \frac{P}{T} =  \ r e^{i \delta} e^{i \phi}  \ \ , \ \  r = \  r'|{\frac{V_{tb}V^*_{tq}}{V_{ub}V^*_{uq}}}|$$ 

 where $r'$ is the "naked" ratio $P/T \ $. It is almost constant over the $\pi \pi$ interval mass, 
but it varies {\it very sharply} in the $\omega$ interval, from 760 MeV $ \ \to \ $ 820 MeV, especially in the
channel $B^0 \to K^{*0} {\rho}^0 {(\omega)} \ $ where it reaches $60\%$. 
 Its variation is almost independent of $ {q^2}/{m^2_b} \ $
but it depends  on $ N^{eff}_c$.
        \begin{center} 
            \begin{tabular}{cccc} 
               \hline 
                \hline
                  Channel & Usual Values & $\omega \ $ Interval\\
               $B^0 \to K^{* 0} \rho^0(\omega)$ & $0.08 \leq {r'}  \leq  0.30$ & 0.60 & \\ 
               \hline  
               $B^+ \to K^{* +} \rho^0(\omega)$ & $0.04 \leq {r'}  \leq  0.05$ &  0.06 &\\ 
                 \hline
               $B^+ \to \rho^{+} \rho^0(\omega)$ & $r' \approx  0.016 $  &  0.05 & \\ 
               \hline  	       
                \hline     
          \end{tabular} 
         \end{center}

 \subsubsection*{ Final Phase-Shift $\delta$ }
 
 The strong phase $\delta$ which is the phase difference between the Penguin and Tree diagrams is the main
ingredient of the {\it absorptive part} of the B decay amplitude. Its physical origin is related to the~:
(i) Intrinsic pahse-shift induced by the Top quark in the Penguin diagram, (ii) the complex Wilson
Coefficients, and (iii) essentially the ${\rho}^0 - \omega$ mixing in the $\pi \pi$ final state interactions.
It depends on the ratio ${q^2}/{{m_b}^2}$ and {\it strongly} on $ N^{eff}_c$. Usually, $\delta$ is almost 
constant in all the $\pi \pi$ invariant mass interval except in the $\omega $ resonance window,  
$ 770 \to  790 $ MeV, where it undergoes a variation of $ 80^{\circ}  \to  100^{\circ} $ in the
$ K^{*0} {\rho}^0 ({\omega})$ channel and a variation of $ 5^{\circ}  \to  25^{\circ} $ in the 
$ K^{*+} {\rho}^0 ({\omega})$ one.
\vskip 0.2cm

$\bullet$ Very interesting physical consequences can be inferred from the exhaustive study of the parameters 
$P/T  \  \mathrm{and} \  \delta $ with the $\pi \pi$ invariant mass. The direct CP asymmetry parameter
is defined according to the relation~: 
\begin{equation}\label{eq14}
a_{CP}^{dir} = {\frac {A^2 -{\bar A}^2}{A^2 +{\bar A}^2}} = \frac{-2 \ r \ {\sin {\delta}} \ {\sin {\phi}}}{1 + r^2
+2 \ r \ {\cos {\delta}} \ {\cos {\phi}}}\ . 
\end{equation}

where $\Phi$ is one of the weak mixing angle deduced from the CKM matrix elements. In the channel 
$B \to {\rho}^0{(\omega)} K^* \ $, angle $\Phi$ is identified with $Arg{[V_{tb} V^{*}_{ts}/V_{ub} V^{*}_{us}]}
 = \gamma \ $; while in the channel $B \to {\rho}^0{(\omega)} \rho $, angle $\Phi$ is given by 
 $ Arg{[V_{tb} V^{*}_{td}/V_{ub} V^{*}_{ud}]} = \   {\beta} + {\gamma} \ = {\Pi} - {\alpha} $.
So, the theoretical knowledge of $r \ \mathrm{and}  \  \delta$ and the experimental measurements 
of $a_{CP}^{dir}$ according to the $ \pi \pi$ invariant mass allow to extract angle(s) $\Phi$ 
from the above equation~(\ref{eq14}). 
\par
These results could be seen as {\it experimental challenges} for the future LHC experiments in the 
field of $B$ physics like LHCB.

\subsubsection*{Recent Experimental Results}

Recently, B factories like BaBar and Belle experiments published interesting results related to the
charmless decays $B \to V_1 V_2$. They both agree on the fact that the longitudinal part of the decay
amplitude is {\it very dominant}, which is one of our essential results. However, these collaborations
do not take into account the process of ${\rho}^0 - {\omega}$ mixing in the estimation of the branching ratios
and the asymmetries. By computing the branching ratios from relation (\ref{eq13}) and comparing them with those
published in ref. \cite{Babar} and \cite{Belle},  we can summarize the main results in the table~\ref{tab1}  
\begin{table}[thb]
\begin{center}
\begin{tabular}{|l|c|c|c|}
\hline
 Channel&Br($\times  {10}^{-6}$)&$f_L = {|H_0|}^2$&$A_{CP}$\\
\hline
${\rho}^0 K^{*+}$(BaBar)&${10.6}^{+3.0}_{-2.6} {\pm} 2.4$&${0.96}^{+0.04}_{-0.15} {\pm}0.04$&${0.20}^{+0.32}_{-0.29}
{\pm} 0.04$\\
\hline
 Our results&$2.3 \to 5.8$&$87\%$&${-6.4\%}  \to   {-22\%}$\\
\hline
\hline
${\rho}^0 {\rho}^+$(BaBar)&${22.5}^{+5.7}_{-5.4} {\pm} 5.8$&${0.97}^{+0.03}_{-0.07} {\pm}0.04$&${-0.19} {\pm}0.23
{\pm} 0.03$\\
\hline
${\rho}^0 {\rho}^+$(Belle)&${31.7}^{+3.8}_{-6.7} {\pm} 7.1$&${0.95} {\pm}0.02 {\pm}0.11$&${0.00} {\pm}0.22
{\pm} 0.03$\\
\hline
Our results&$11.0 \to 20.0$&$90\%$&${-8.5\%}  \to   {-10\%}$\\
\hline
\hline
\end{tabular}
\end{center}
\caption{Predicted results compared to experimental data from BaBar and Belle}
\label{tab1}
\end{table}

\section {Conclusion and Perspectives}

Helicity formalism has been used very successfully for a full computation and numerical simulations of 
the channels $B \to {\pi}^+ {\pi}^- V$  with  $V = \ K^{*0}, K^{*{\pm}}, {\rho}^{\pm}$. Naive factorization
is very useful for weak hadronic B decays despite its theoretical uncertainties. Furthermore, interesting
results have been obtained like~: (i) The important role of the {\it form factor models}, 
 (ii) The longitudinal polarization is largely dominant, 
whatever the form factor model.
(iii) The ${{\rho}^0}- {\omega}$ mixing is the main ingredient in 
the {\it enhancement} of the direct $CP$ violation.
 (iv) A new way to look for direct $CP$ Violation is found and it can help to 
develop new methods  for measuring the  angles 
{\bf ${\gamma}  \  {\rm and} \  {\alpha} $}. \\
What remains to be done is to cross-check these predictions with experimental data coming soon from the
LHC exoeriments. \\
{\underline{Acknowledgments}}~: Z.J.A. is very indebted to the organizers of the QFTHEP04 conference 
which was held in this historical and marvellous city of Sant-Petersburg.


\begin{figure}[thb]
\centering\includegraphics[height=10.0cm,clip=true]{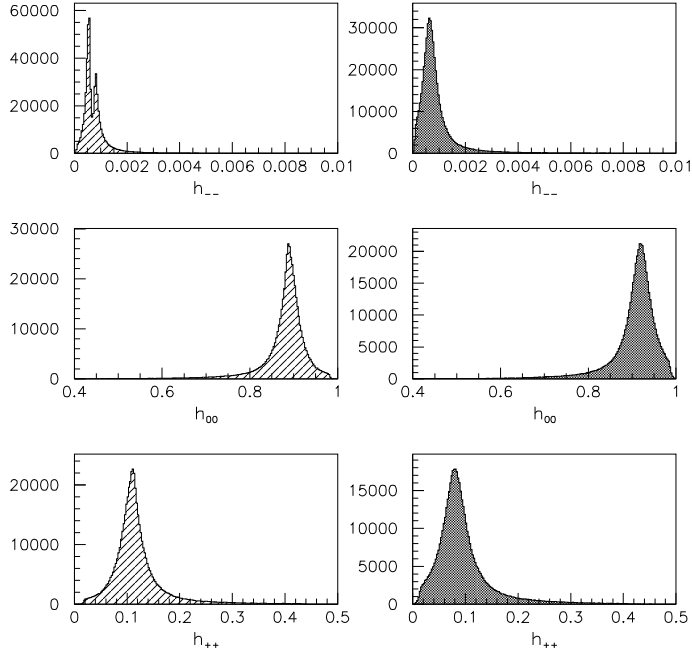}
\caption{Spectrum of $ h_{--}, h_{00}, h_{++} \ $~:
$B^0 \rightarrow \rho^{0}(\omega) K^{* 0} \ $ (left)
$ \ B^+ \rightarrow \rho^{0}(\omega) \rho^+$ (right) }
\end{figure}
\begin{figure}[thb]
\centering\includegraphics[height=10.0cm,clip=true]{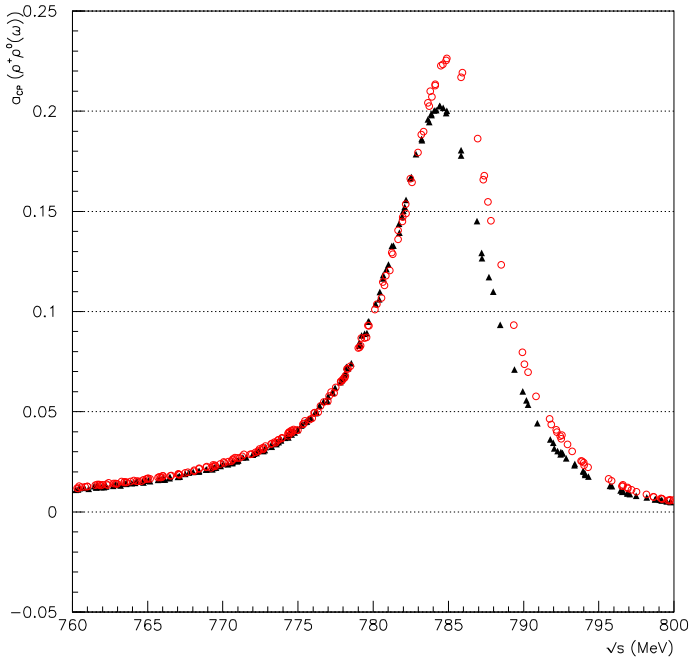}
\caption{ $CP$-violating asymmetry parameter,  $a_{CP}(m)$, as a function of the
 ${\pi}^+ {\pi}^- $ 
invariant mass in the vicinity of the $\omega$ mass region for the channel $B^+ \rightarrow \rho^{0}(\omega)
  \rho^+ \ $ for two different form factor models.}
\end{figure}

%


\begin{thebibliography}{99}
\bibitem{Dunietzetal}
I. Dunietz {\it et al}, Phys. Rev. {\bf D43}  (1991) 2193.
\bibitem{Ziadetal}
Z.J. Ajaltouni {\it et al}, Eur.Phys.J. {\bf C 29}, 215-233 (2003).
\bibitem{Quinn}
H. Quinn, {\it "Hadronic effects in two-body B decays"}. Lectures at SLAC Summer Institute (1999).
\bibitem{Burasetal} A.J. Buras, Lect. Notes Phys. {\bf 558} (2000) 65, also in `Recent Developments in Quantum Field 
Theory',
Springer Verlag, edited by P. Breitenlohner, D. Maison and J. Wess (Springer-Verleg, Berlin, in press),
 hep-ph/9901409;
 R. Fleischer, Int. J. Mod. Phys. {\bf A12} (1997) 2459, Z. Phys. {\bf C62} (1994) 81, 
Z. Phys. {\bf C58} (1993) 483.
\bibitem{BSW} 
M. Bauer, B. Stech and M. Wirbel, Z. Phys. {\bf C34} (1987) 103; 
M. Wirbel, B. Stech and M. Bauer, Z. Phys.  {\bf C29} (1985) 637.
\bibitem{Rimbault}
C. Rimbault, PhD Thesis, DU1492 ''{\it Etude de la violation directe de $CP$ dans la 
d\'esintegration du m\'eson $B$ en deux m\'esons vecteurs non charm\'es. Analyse
du canal $K^{*0} \rho^{0}(\omega)$ dans le cadre de l'experience LHCb.}'' Universit\'e Blaise Pascal-Clermont
II (F\'evrier 2004) \\
O. Leitner, PhD Thesis, ''{\it Direct $CP$ violation in $B$ decays including $\rho^{0}-\omega$ mixing 
and covariant light-front dynamics}''.
\bibitem{BSS}
M.Bander, D.Silverman, A.Soni, Phys.Rev.Let. {\bf 43} (1979) 242
\bibitem{Babar}
B. Aubert {\it et al} (BaBar collaboration), "Rates, Polarizations and asymmetries in Charmless Vector-Vector
 B Meson Decays", Phys.Rev.Let. {\bf 91} (2003), 171802 and Phys.Rev. {\bf D69}, 031102(R) (2004).
\bibitem{Belle} 
J. Zhang {\it et al} (Belle collaboration), "Observation of $B^{\pm} \to {\rho}^{\pm} {\rho}^0$ "  \\
 Phys.Rev.Let. {\bf 91} (2003), 221801
\bibitem{Thomas}
O. Leitner, X.-H. Guo, A.W.Thomas, Phys. Rev. {\bf D63} (2001) 056012.
\bibitem{Leitner}
O.Leitner, X.Guo, A.W.Thomas, Phys.Rev. {\bf D66} (2002), 096008
\end{thebibliography}
\end{document}